\documentclass{aa}

\usepackage{txfonts}
\usepackage{natbib}
\usepackage{amsbsy}
\usepackage{graphicx}
\usepackage{aalongtable}

\bibpunct{(}{)}{;}{a}{}{,}     

\newcommand{\micron}{$\mu$m}

\newcommand{\el}[1]{\mathrm{#1}}

\newcommand{\ch}[3]{\el{C}_{#1}^{}\el{H}_{#2}^{#3}}

\newcommand{\ccps}{$\el{cm}^3~\el{s}^{-1}$}

\newcommand{\scit}[2]{$#1\times10^{#2}$}
\newcommand{\scim}[2]{#1\times10^{#2}}

\newcommand{\pcc}{$\el{cm}^{-3}$}

\newcommand{\eq}[1]{Eq.~(\ref{eq:#1})}
\newcommand{\eqq}[1]{Equation (\ref{eq:#1})}
\newcommand{\fig}[1]{Fig.~\ref{fig:#1}}
\newcommand{\figg}[1]{Figure \ref{fig:#1}}
\newcommand{\tb}[1]{Table~\ref{tb:#1}}
\newcommand{\nc}{N_\el{C}^{}}
\newcommand{\nh}{N_\el{H}^{}}
\newcommand{\nho}{N_\el{H}^\circ}

\begin{document}

\title{PAH chemistry and IR emission from circumstellar disks}

\author{
R. Visser \inst{1}
 \and V.C. Geers \inst{1}
 \and C.P. Dullemond \inst{2}
 \and J.-C. Augereau \inst{1,3}
 \and K.M. Pontoppidan \inst{4,5}
 \and E.F. van Dishoeck \inst{1}
}

\institute{Leiden Observatory, Leiden University, P.O. Box 9513, 2300 RA Leiden, The Netherlands
 \and Max-Planck-Institut f\"{u}r Astronomie, Koenigstuhl 17, 69117 Heidelberg, Germany
 \and Laboratoire d'Astrophysique de l'Observatoire de Grenoble, B.P. 53, 38041 Grenoble Cedex 9, France
 \and Division of GPS, Mail Code 150-21, California Institute of Technology, Pasadena, CA 91125, USA
 \and Hubble Fellow
}

\authorrunning{Visser et al.}

\date{Received $<$date$>$ / Accepted $<$date$>$}


\abstract
{} 
{The chemistry of, and infrared (IR) emission from, polycyclic aromatic hydrocarbons (PAHs) in disks around Herbig Ae/Be and T Tauri stars are investigated. PAHs can exist in different charge states and they can bear different numbers of hydrogen atoms. The equilibrium (steady-state) distribution over all possible charge/hydrogenation states depends on the size and shape of the PAHs and on the physical properties of the star and surrounding disk.} 
{A chemistry model is created to calculate the equilibrium charge/hydrogenation distribution. Destruction of PAHs by ultraviolet (UV) photons, possibly in multi-photon absorption events, is taken into account. The chemistry model is coupled to a radiative transfer code to provide the physical parameters and to combine the PAH emission with the spectral energy distribution (SED) from the star+disk system.} 
{Normally hydrogenated PAHs in Herbig Ae/Be disks account for most of the observed PAH emission, with neutral and positively ionized species contributing in roughly equal amounts. Close to the midplane, the PAHs are more strongly hydrogenated and negatively ionized, but these species do not contribute to the overall emission because of the low UV/optical flux deep inside the disk. PAHs of $50$ carbon atoms are destroyed out to $100$ AU in the disk's surface layer, and the resulting spatial extent of the emission does not agree well with observations. Rather, PAHs of about $100$ carbon atoms or more are predicted to cause most of the observed emission. The emission is extended on a scale similar to that of the size of the disk, with the short-wavelength features less extended than the long-wavelength features. For similar wavelengths, the continuum emission is less extended than the PAH emission. Furthermore, the emission from T Tauri disks is much weaker and concentrated more towards the central star than that from Herbig Ae/Be disks. Positively ionized PAHs are predicted to be largely absent in T Tauri disks because of the weaker radiation field.} 
{} 


\maketitle


\section{Introduction}
\label{sec:intro}
Polycyclic aromatic hydrocarbons \citep[PAHs;][]{leger84a,allamandola89a} are ubiquitous in space and are seen in emission from a wide variety of sources, including the diffuse interstellar medium, photon-dominated regions, circumstellar envelopes, and (proto)planetary nebulae \citep[and references therein]{peeters04a}. The PAHs in these sources are electronically excited by ultraviolet (UV) photons. Following internal conversion to a high vibrational level of the electronic ground state, they cool by emission in the C--H and C--C stretching and bending modes at $3.3$, $6.2$, $7.7$, $8.6$, $11.3$, $12.8$ and $16.4$ \micron.

Using the Infrared Space Observatory \citep[ISO;][]{kessler96a}, the Spitzer Space Telescope \citep{werner04a,houck04a} and various ground-based telescopes, PAH features have also been observed in disks around Herbig Ae/Be and T Tauri stars \citep{vankerckhoven00a,hony01a,peeters02a,przygodda03a,vanboekel04b,acke04a,geers06a}. Spatially resolved observations confirm that the emission comes from regions whose size is consistent with that of a circumstellar disk \citep{vanboekel04a,geers05a,habart06a}. Because of the optical and/or UV radiation required to excite the PAHs, their emission is thought to come mostly from the surface layers of the disks \citep{habart04a}. \citet{acke04a} showed that PAH emission is generally stronger from flared disks \citep{meeus01a,dominik03a} than from flat or self-shadowed disks.

Although the presence of such large molecules in disks and other astronomical environments is intrinsically interesting, it is also important to study PAHs for other reasons. They are a good diagnostic of the stellar radiation field and can be used to trace small dust particles in the surface layers of disks, both near the center and further out \citep{habart04a}. In addition, they are strongly involved in the physical and chemical processes in disks. For instance, photoionization of PAHs produces energetic electrons, which are a major heating source of the gas \citep{bakes94a,kamp04a,jonkheid04a}. The absorption of UV radiation by PAHs in the surface layers influences radiation-driven processes closer to the midplane. Charge transfer of $\el{C}^+$ with neutral and negatively charged PAHs affects the carbon chemistry. Finally, \citet{habart04c} proposed PAHs as an important site of H$_2$ formation in photon-dominated regions, a process which is also important in disk chemistry \citep{jonkheid06b,jonkheid06a}. Although this process is probably more efficient on grains and very large PAHs than it is on PAHs of up to $100$ carbon atoms, the latter may play an important role if the grains have grown to large sizes.

Many PAHs and related species that are originally present in the parent molecular cloud, are able to survive the star formation process and eventually end up on planetary bodies \citep{allamandola03a}. They add to the richness of the organochemical ``broth'' on planets in habitable zones \citep{kasting93a}, from which life may originate. Further enrichment is believed to come from the impact of comets. PAHs have now been detected in cometary material during the Deep Impact mission \citep{lisse06a} and returned to Earth by the Stardust mission \citep{sandford06a}. The icy grains that constitute comets also contain a variety of other molecules \citep{ehrenfreund03a}. Radiation-induced chemical reactions between frozen-out PAHs and these molecules lead to a large variety of complex species, including some that are found in life on Earth \citep{bernstein99a,ehrenfreund06a}. These possibilities are another reason why it is important to study the presence and chemistry of PAHs in disks.

The chemistry of PAHs in an astronomical context has been studied with increasingly complex and accurate models in the last two decades \citep{omont86a,lepp88a,bakes94a,salama96a,dartois97a,vuong00a,lepage01a,lepage03a,weingartner01a,bakes01a,bakes01b}; however, none of these were specifically targeted at PAHs in circumstellar disks. Disk chemistry models that do include PAHs only treat them in a very simple manner \citep[e.g.][]{jonkheid04a,habart04a}. In this paper, an extensive PAH chemistry model is coupled to a radiative transfer model for circumstellar disks \citep[Dullemond et al. in prep.]{dullemond04a,geers06a}. The chemistry part includes ionization (photoelectric emission), electron recombination and attachment, photodissociation with loss of hydrogen and/or carbon, and hydrogen addition. Infrared (IR) emission from the PAHs is calculated taking multi-photon excitation into account, and added to the spectral energy distribution (SED) of the star+disk system. The model can in principle also be used to examine PAH chemistry and emission in other astronomical environments.

We will present the chemistry model in Section \ref{sec:pahmodel}, followed by a brief review of the radiative transfer model in Section \ref{sec:diskmodel}. The results are discussed in Section \ref{sec:results} and our conclusions are summarized in Section \ref{sec:con}.


\section{PAH model}
\label{sec:pahmodel}

The chemistry part of our model is a combination of the models developed by \citet[hereafter LPSB01]{lepage01a} and \citet[hereafter WD01]{weingartner01a}. Photodissociation is treated according to \citet{leger89a}. Where possible, theoretical rates are compared to recent experimental data. Our model employs the new PAH cross sections of \citet[hereafter DL06]{draine06a}, which are an update of \citet{li01a} based on experimental data \citep{mattioda05a,mattioda05b} and IR observations \citep[e.g.][]{werner04b,smith04b}. In this section, we present the main characteristics of our model.


\subsection{Characterization of PAHs}
\label{subsec:pahchar}
The PAHs in our model are characterized by their number of carbon atoms, $\nc$. A PAH bears the normal number of hydrogen atoms, $\nho$, when one hydrogen atom is attached to each peripheral carbon atom bonded to exactly two other carbon atoms (e.g. $12$ for coronene, $\ch{24}{12}{}$). The ratio between $\nho$ and $\nc$ is taken as (e.g. DL06):
\begin{equation}
\label{eq:h/c}
f^\circ = \nho/\nc = 
\left\{\begin{array}{ll} 
0.5                    & \nc\leq25\,, \\
0.5\sqrt{25/\nc}\qquad & 25<\nc\leq100\,, \\
0.25                   & \nc>100\,.
\end{array}\right.
\end{equation}
This formula produces values appropriate for compact (pericondensed) PAHs. Elongated (catacondensed) PAHs will have a higher hydrogen coverage; however, they are believed to be less stable and to convert into a more compact geometry \citep{wang97a,dartois97a}, so only compact PAHs are assumed to be present.

Every carbon atom that bears one hydrogen atom in the normal case is assumed to be able to bear two in extreme conditions, so each PAH can exist in $2\nho+1$ possible hydrogenation states ($0\le\nh\le2\nho$). Furthermore, each PAH can exist in two or more charge states $Z$. The maximum and minimum attainable charge depend on the radiation field and the PAH's ionization and autoionization potentials \citep[WD01;][]{bakes94a}. The number of accessible charge states increases with PAH size.

For a number of properties, the radius of the PAH is more important than $\nc$. The PAHs of interest are assumed to be spherically symmetric and, following e.g. WD01 and \citet{draine01a}, are assigned an effective radius $a$,
\begin{equation}
\label{eq:radius}
a = \left(\frac{\nc}{0.468}\right)^{1/3}\,,
\end{equation}
where $a$ is in \AA. This equation does not give the actual geometric radius; rather, it gives the radius of a pure graphite sphere containing the same number of carbon atoms. Equations (\ref{eq:h/c}) and (\ref{eq:radius}) do not apply to very small PAHs ($\nc\la20$), but like irregularly shaped PAHs, they are assumed not to be abundant enough to contribute to the emission (Section \ref{subsec:growth}).


\subsection{Photoprocesses}
\label{subsec:pp}
The absorption of a UV or visible photon of sufficient energy by a PAH causes either the emission of an electron (ionization or detachment) or a transition to an excited electronic and vibrational state. In the second case, internal conversion and fluorescence rapidly bring the molecule to a high vibrational level of the ground electronic state. From here, several processes can take place \citep{leger88a,leger89a}: (1) dissociation with loss of atomic or molecular hydrogen; (2) dissociation with loss of a carbon-bearing fragment; or (3) cooling by infrared emission. The cooling rate constant for a PAH with internal energy $E_\el{int}$ is given by \citet{li03b}:
\begin{equation}
\label{eq:kcool}
k_\el{cool} = \int_{912\el{\AA}}^{\infty}\frac{4\pi{}B_\lambda(T[E_\el{int}])\sigma_\el{abs}}{hc/\lambda}\el{d}\lambda\,,
\end{equation}
with $B_\lambda(T[E_\el{int}])$ the Planck function at the PAH's vibrational temperature, $T[E_\el{int}]$ \citep{draine01a}. The cross sections, $\sigma_\el{abs}$, will be treated in Section \ref{subsec:cross}.

The yield of a single dissociation process $i$, $Y_i$, depends on the rate constants, $k$, of all possible processes:
\begin{equation}
\label{eq:pyld}
Y_i = \frac{k_i}{\sum_j k_j}\,.
\end{equation}
To determine the rate $\Gamma_i$ of dissociation process $i$, the yield is multiplied by the absorption rate, taking into account the possibility of electron emission, and integrated over the relevant energy range:
\begin{equation}
\label{eq:prate}
\Gamma_i = \int_{0}^{\infty}(1-Y_\el{em})Y_i\sigma_\el{abs}N_\el{ph}\el{d}E\,,
\end{equation}
where $Y_\el{em}$ is the photoelectric emission yield (\eq{yem}) and $N_\el{ph}$ is the number of photons in units of $\el{cm}^{-2}~\el{s}^{-1}~\el{erg}^{-1}$. The radiation field is treated explicitly at every point in the disk in an axisymmetric $3$-D geometry, i.e., the wavelength dependence is taken into account as well as the magnitude.


Photoelectric emission is the ejection of an electron from a PAH due to the absorption of a UV photon. The electron can come either from the valence band (photoionization; possible for all charge states) or from an energy level above the valence band (photodetachment; only for negatively charged PAHs). The photoelectric emission rate is given by WD01 as:
\begin{equation}
\label{eq:per1}
\Gamma_{\el{em}} = \int_{h\nu_\el{ion}}^{\infty}Y_\el{ion}\sigma_\el{abs}N_\el{ph}\el{d}E+\int_{h\nu_\el{det}}^{\infty}Y_\el{det}\sigma_\el{det}N_\el{ph}\el{d}E\,,
\end{equation}
where $h$ is Planck's constant. Photoelectric emission will only occur when the photon energy exceeds the threshold of $h\nu_\el{ion}$ or $h\nu_\el{det}$. The photodetachment yield, $Y_\el{det}$, is taken to be unity: every absorption of a photon with $h\nu>h\nu_\el{det}$ leads to the ejection of an electron. The photoionization yield, $Y_\el{ion}$, has a value between $0.1$ and $1$ for $h\nu\ga8$ eV and drops rapidly for lower energies. For photons with $h\nu>h\nu_\el{ion}$, the photodetachment cross section, $\sigma_\el{det}$, is about two orders of magnitude lower for $\nc=50$ than the ionization cross section, which is assumed equal to $\sigma_\el{abs}$.

In order to derive $Y_\el{em}$ as used in \eq{prate}, \eq{per1} is rewritten:
\begin{equation}
\label{eq:per2}
\Gamma_{\el{em}} = \int_{0}^{\infty}\left(Y_\el{ion}+\frac{\sigma_\el{det}}{\sigma_\el{abs}}Y_\el{det}\right)\sigma_\el{abs}N_\el{ph}\el{d}E\,.
\end{equation}
The quantity between brackets is the electron emission yield, $Y_\el{em}$:
\begin{equation}
\label{eq:yem}
Y_\el{em} = Y_\el{ion}+\frac{\sigma_\el{det}}{\sigma_\el{abs}}Y_\el{det}\,,
\end{equation}
taking $Y_\el{ion}$ and $Y_\el{det}$ to be zero for photon energies less than $h\nu_\el{ion}$ and $h\nu_\el{det}$, respectively. Furthermore, $Y_\el{em}$ is not allowed to exceed unity: each absorbed photon can eject only one electron.


If no photoelectric emission takes place, the PAH can undergo dissociation with loss of carbon or hydrogen. Several theoretical schemes exist to calculate the dissociation rates. LPSB01 employed the Rice-Ramsperger-Kassel-Marcus quasi-equilibrium theory (RRKM-QET) and obtained hydrogen loss rates close to those determined experimentally for benzene ($\ch{6}{6}{}$), naphthalene ($\ch{10}{8}{}$) and anthracene ($\ch{14}{10}{}$). \citet{leger89a} investigated the loss of carbon as well as hydrogen, using an inverse Laplace transform of the Arrhenius law \citep{forst72a} to determine the rates. In this method, the rate constant is zero when the PAH's internal energy, $E_\el{int}$, originating from one or more UV photons, is less than the critical energy, $E_0$, for a particular loss channel. When $E_\el{int}$ exceeds $E_0$,
\begin{equation}
\label{eq:kdiss}
k_\el{diss,X} = A_\el{X}\frac{\rho(E_\el{int}-E_{0,\el{X}})}{\rho(E_\el{int})}\,,
\end{equation}
where $A_\el{X}$ is the pre-exponential Arrhenius factor for channel X and $\rho(E)$ is the density of vibrational states at energy $E$. \citet{leger89a} considered the loss of H, C, C$_2$ and C$_3$; LPSB01 also took H$_2$ loss into account for PAHs with $\nh>\nho$. Our model uses the method of \citet{leger89a} for all loss channels, with values for $A$ and $E_0$ given in \tb{disspar}.

The intensity of the UV radiation field inside the disk is characterized in our model by $G_0'=u_\el{UV}^{}/u_\el{UV}^\el{Hab}$, where
\begin{equation}
\label{eq:g_0}
u_\el{UV}^{} = \int_{6\ \el{eV}}^{13.6\ \el{eV}}u_\nu\el{d}\nu = \int_{6\ \el{eV}}^{13.6\ \el{eV}}(h/c)h\nu{}N_\el{ph}{d}\nu
\end{equation}
and $u_\el{UV}^\el{Hab}=\scim{5.33}{-14}$ erg~\pcc{} is the energy density in the mean interstellar radiation field \citep{habing68a}. However, the radiation field inside the disk does not have the same spectral shape as the interstellar radiation field, so $G_0'$ is used instead of $G_0^{}$ to denote its integrated intensity. The exact UV field at every point and every wavelength is calculated by a Monte Carlo code, which follows the photons from the star into the disk in an axisymmetric $3$D geometry (see also Section \ref{subsec:code}). As a result, regions of high optical depth may not receive enough photons between $6$ and $13.6$ eV to calculate $G_0'$. In those cases, the lower limit of the integration range is extended in $1$-eV steps and the energy density of the Habing field is recalculated accordingly. If $G_0'$ still cannot be calculated between $1$ and $13.6$ eV, it is set to a value of $10^{-6}$.

In the strong radiation fields present in circumstellar disks (up to $G_0'=10^{10}$ in the inner disk around a Herbig Ae/Be star), multiple photons are absorbed by a PAH before it can cool through emission of IR radiation. These multi-photon events result in higher dissociation rates than given by Eqs.~(\ref{eq:prate}) and (\ref{eq:kdiss}). For example, a PAH of $50$ carbon atoms is photodestroyed in a radiation field of $G_0'\approx10^5$ if multi-photon events are allowed, while $G_0'\approx10^{14}$ is required in a pure single-photon treatment (Section \ref{subsec:growth}). However, the multi-photon treatment is much more computationally demanding, hence our model is limited to single-photon treatments for all processes but photodestruction; the latter will be discussed in detail in Section \ref{subsec:growth}. The errors introduced by not including a full multi-photon treatment will be discussed in Section \ref{subsec:sens}.

\begin{table}
\caption{Arrhenius parameters for the loss of carbon and hydrogen fragments from PAHs.$^{\mathrm{a}}$}
\label{tb:disspar}
\centering
\begin{tabular}{ccccc}
\hline\hline
Fragment          & $\nh$     & $Z$    & $E_0$ (eV)       & $A$ (s$^{-1}$) \\
\hline
C$_{\phantom{1}}$ & all       & all    & $7.37$           & $\scim{6.2}{15}$ \\
C$_2$             & all       & all    & $8.49$           & $\scim{3.5}{17}$ \\
C$_3$             & all       & all    & $7.97$           & $\scim{1.5}{18}$ \\
H$_{\phantom{1}}$ & $\le\nho$ & all    & $4.65$           & $\scim{1.5}{15}$ \\
H$_{\phantom{1}}$ & $>\nho$   & $\le0$ & $1.1\phantom{0}$ & $\scim{\phantom{0.}4}{13}$ \\
H$_{\phantom{1}}$ & $>\nho$   & $>0$   & $2.8\phantom{0}$ & $\scim{\phantom{0.}1}{14}$ \\
H$_2$             & $>\nho$   & $\le0$ & $1.5\phantom{0}$ & $\scim{\phantom{0.}4}{13}$ \\
H$_2$             & $>\nho$   & $>0$   & $3.1\phantom{0}$ & $\scim{\phantom{0.}1}{14}$ \\
\hline
\end{tabular}
\begin{list}{}{}
\item[$^{\mathrm{a}}$] $E_0$ and $A$ for C, C$_2$ and C$_3$ are from \citet{leger89a}. $E_0$ for H and H$_2$ is based on the RRKM-QET parameters from \citet{lepage01a} and modified slightly to obtain a better match with their rates. $A$ for H is modified from \citet{leger89a} and used also for H$_2$.
\end{list}
\end{table}


As long as the internal energy of the PAH exceeds $E_0$ for any of the loss channels, there is competition between dissociation and radiative stabilization. Typically, the emission of a single IR photon is not sufficient to bring $E_\el{int}$ below $E_0$. If the emitted IR photons are assumed to have an average energy $q_\el{IR}$ of $0.18$ eV ($6.9$ \micron; LPSB01), a total of $n=(E_\el{int}-E_0)/q_\el{IR}$ photons are required to stabilize the PAH. The $i$th photon in this cooling process is emitted at an approximate rate \citep{herbst91a}
\begin{equation}
\label{eq:kradi}
k_{\el{rad},i} = 73\frac{[E_\el{int}-(i-1)q_\el{IR}]^{1.5}}{s^{0.5}}\,,
\end{equation}
with $E_\el{int}$ in eV and $s=3(\nc+\nh)-6$ the number of vibrational degrees of freedom. The competition between dissociation and IR emission occurs for every intermediate state. Hence, the total radiative stabilization rate is \citep{herbst99a}
\begin{equation}
\label{eq:krad}
k_\el{rad} = k_{\el{rad},1}\prod_{i=1}^n\frac{k_{\el{rad},i}}{k_{\el{rad},i}+k_\el{diss,H}+k_\el{diss,H_2}+k_\el{diss,C}}\,,
\end{equation}
where $k_\el{diss,C}$ denotes the sum of all three carbon loss channels. Note that Eqs.~(\ref{eq:kradi}) and (\ref{eq:krad}) are used only in the chemical part of the code, where the details of the cooling rate function are not important (LPSB01). The radiative transfer part employs the more accurate \eq{kcool}.

Alternative stabilization pathways such as inverse internal conversion, inverse fluorescence and Poincar\'e fluorescence \citep{leach87a,leger88a} are probably of minor importance. Their effect can be approximated by increasing $q_\el{IR}$, thus creating an ``effective'' $k_\el{rad}$ that is somewhat larger than the ``old'' $k_\el{rad}$. However, this will have no discernable effect on the rates from \eq{prate}, so these alternative processes are ignored altogether.


\subsection{Absorption cross sections}
\label{subsec:cross}
\citet{li01a} performed a thorough examination of the absorption cross sections, $\sigma_\el{abs}$, of PAHs across a large range of wavelengths. Following new experimental data \citep{mattioda05a,mattioda05b} and IR observations \citep[e.g.][]{werner04b,smith04b}, an updated model was published in DL06, providing a set of equations that can readily be used in our model. The cross sections consist of a continuum contribution that decreases towards longer wavelengths, superposed onto which are a number of Drude profiles to account for the $\sigma$--$\sigma^\ast$ transition at $72.2$ nm, the $\pi$--$\pi^\ast$ transition at $217.5$ nm, the C--H stretching mode at $3.3$ \micron, the C--C stretching modes at $6.2$ and $7.7$ \micron, the C--H in-plane bending mode at $8.6$ \micron, and the C--H out-of-plane bending mode at $11.2$--$11.3$ \micron. Some of these primary features are split into two or three subfeatures and several minor features are included in the $5$--$20$ \micron{} range to give a better agreement with recent observations (DL06). The additional absorption for ions in the near IR measured by \citet{mattioda05a,mattioda05b} is included as a continuum term and three Drude profiles.

The absorption properties of a PAH depend on its charge. Neutral PAHs have $6.2$, $7.7$ and $8.6$ \micron{} features that are a factor of a few weaker than do cations. The $3.3$ \micron{} band strength increases from cations to neutrals \citep{bauschlicher02a,hudgins01a,hudgins00a,langhoff96a}. The other features have similar intensities for each charge state. While the cation/neutral band ratios at $6.2$, $7.7$ and $8.6$ \micron{} are mostly independent of size, those at $3.3$ \micron{} decrease for larger PAHs. DL06 provide integrated band strengths, $\sigma_\el{int}^{}$, for neutral and ionized PAHs, but they do not account for the size dependence of this feature in cations. Rather than taking $\sigma_\el{int,3.3}^{Z=1} = 0.227\sigma_\el{int,3.3}^{Z=0}$ as DL06 do, our model uses
\begin{equation}
\label{eq:sigintcat}
\sigma_\el{int,3.3}^{Z=1} = \sigma_\el{int,3.3}^{Z=0}\times\left(1+\frac{41}{\nc-14}\right)^{-1}\,.
\end{equation}
This way, the ionized/neutral band ratio approaches unity for large PAHs. The theoretical ratios for $\nc=24$, $54$ and $96$ from \citet{bauschlicher02a} are well reproduced by \eq{sigintcat}.

LPSB01 based their opacities on experiments by \citet{joblin92a} on PAHs in soot extracts. They fitted the cross section of an extract with an average PAH mass of $365$ amu (corresponding to $\nc\approx30$) with a collection of Gaussian curves, obtaining practically the same values for $\lambda\la0.25$ \micron{} ($E\ga5$ eV) as DL06. Between $0.25$ and $0.5$ \micron{}, the integrated cross section of LPSB01 is about $2.5$ times stronger. Since LPSB01 were only interested in the UV opacities, their cross section throughout the rest of the visible and all of the infrared is zero. For our purposes, the different opacities in the $0.25$--$0.5$ \micron{} range will only affect the hydrogen dissociation rates, and, as will be shown in Section \ref{subsec:sens}, the effects are negligible.

The details of the infrared cross sections of negatively charged PAHs are not well understood, with very different values to be found in the literature \citep{bauschlicher00a,langhoff96a}. However, this is not a problem in this work. Because of the small electron affinities, photodetachment is a very efficient process and absorption of a photon by a PAH anion leads to ejection of an electron rather than substantial emission in the infrared. Rapid electron attachment retrieves the original anion before the transient neutral PAH can absorb a second photon. Hence, if anions are present in steady state, they are excluded altogether when the infrared emission is calculated.

Large PAHs can carry a double or triple positive charge in those regions of the disk where the optical/UV field is strong and the electron density is low. The differences between the cross sections of singly charged and neutral PAHs generally increase when going to multiply charged PAHs \citep{bauschlicher00a,bakes01a}. However, no cross sections for multiply charged species exist that can be directly used in our model, so the cross sections for singly charged species are used instead. Since multiply charged species only constitute a very small part of the PAH population (Section \ref{subsec:pahdis}), this approximation is not expected to cause major errors.


\subsection{Electron recombination and attachment}
\label{subsec:elatt}
Free electrons inside the disk can recombine with PAH cations or attach to PAH neutrals and anions; both processes will be referred to as electron attachment. The electron attachment rate, $\Gamma_\el{ea}=\alpha_\el{ea}n_\el{e}$, with $n_\el{e}$ the electron number density, depends on the frequency of collisions between electrons and PAHs and on the probability that a colliding electron sticks, as expressed by WD01:
\begin{equation}
\label{eq:ear}
\alpha_\el{ea} = s_\el{e}\left(\frac{8kT}{\pi{}m_\el{e}}\right)^{1/2}\pi{}a^2\widetilde{J}\left(\tau=\frac{4\pi\varepsilon_0akT}{e^2},Z\right)\,,
\end{equation}
where $k$ is Boltzmann's constant, $m_\el{e}$ is the electron mass, $\varepsilon_0$ is the permittivity of vacuum, $T$ is the gas temperature, $e$ is the electron charge and $\tau$ is the so-called reduced temperature.

The sticking coefficient, $s_\el{e}$, has a maximum value of $0.5$ to allow for the possibility of elastic scattering. The electrons that do not scatter elastically, have to be retained by the PAH before their momentum carries them back to infinity. The probability of retention is approximately $1-e^{-a/l_e}$, where $l_e\approx10$ \AA{} can be considered the mean free path of the electron inside the molecule. WD01 include an additional factor $1/(1+e^{20-\nc})$ for PAH neutrals and anions to take into account the possibility that the PAH will be dissociated by the electron's excess energy, but this is only important for PAHs smaller than $24$ carbon atoms (see also LPSB01) and they are assumed not to be present in the disk (Section \ref{subsec:growth}). Thus, our model uses
\begin{equation}
\label{eq:se}
s_\el{e} = 0.5(1-e^{-a/l_e})
\end{equation}
for all PAHs with a charge $Z>Z_\el{min}$.

Expressions for $\widetilde{J}$, a factor that takes into account the charge and size of the PAH and the temperature of the gas, can be found in \citet{draine87a}. For neutral PAHs, $\widetilde{J}\propto{}T^{-1/2}$, so the attachment rate does not depend significantly on the temperature \citep{bakes94a}. $\widetilde{J}$ changes as $T^{-1}$ for PAH cations in the PAH size and gas temperature regimes of interest, leading to $\Gamma_\el{ea}\propto{}T^{-1/2}$. Finally, for PAH anions, both $\widetilde{J}$ and the attachment rate vary as $e^{-1/T}$.

Recombination rates at $300$ K for PAH cations of up to sixteen carbon atoms have been determined experimentally \citep{biennier06a,novotny05a,hassouna03a,rebrionrowe03a,abouelaziz93a}. The experimental and theoretical rates agree to within a factor of $\sim2$ for these small species, except for naphthalene ($\ch{10}{8}{}$), where the theoretical value is larger by a factor of $\sim7$. Experimental data on larger PAHs are needed to ascertain the accuracy of \eq{ear} for the sizes used in our model.

Experimental data on electron attachment to PAH neutrals are scarce. Rates for anthracene ($\ch{14}{10}{}$) have been reported from $\scim{9}{-10}$ to $\scim{4.5}{-9}$ \ccps{} \citep{moustefaoui98a,tobita92a}, which do not depend greatly on temperature. \eqq{ear} predicts a rate of $\sim\scim{2}{-10}$ \ccps{}, about an order of magnitude lower, for temperatures between $10$ and $1000$ K\@. For pyrene ($\ch{16}{10}{}$), however, the theoretical value, $\sim\scim{2}{-9}$ \ccps, is an order of magnitude higher than the values found experimentally, $2$ and $\scim{4.2}{-10}$ \ccps{} \citep{tobita92a}. Hence, we can only estimate that \eq{ear} is accurate to within about an order of magnitude. No experimental data are available on electron attachment to negatively charged PAHs or on electron recombination with PAHs carrying a multiple positive charge, so the uncertainties in the theoretical rates for these reactions are at least as large.


\subsection{Hydrogen addition}
\label{subsec:hydroadd}
The addition of atomic hydrogen to neutral and ionized PAHs is an exothermic process requiring little or no activation energy \citep{hirama04a,bauschlicher98a}. The temperature-independent rates from LPSB01 for addition to cations are used, which are based primarily on experimental work by \citet{snow98a}. Addition to neutrals is about two orders of magnitude slower for benzene \citep{triebert98a,mebel97a} and has not been measured for larger PAHs. No rates are known for the addition of hydrogen to anions or to cations with $Z>1$. We take $k_\el{add,H}^{Z<1} = 10^{-2}k_\el{add,H}^{Z=1}$ and $k_\el{add,H}^{Z>1} = k_\el{add,H}^{Z=1}$. The rate depends on $\nh$ as described by LPSB01.

It is assumed that molecular hydrogen can only attach to PAHs with $\nh<\nho$. Again, the rate for addition to cations from LPSB01 is used and divided by $100$ for the neutral and anion rates.

For naphthalene and larger PAHs, addition of atomic hydrogen (PAH + H $\to$ PAH$_{+1}$) is assumed to be much faster than the bimolecular abstraction channel (PAH + H $\to$ PAH$_{-1}$ + H$_{2}$) \citep{herbst99a}, so the latter is not included in our model.


\subsection{PAH growth and destruction}
\label{subsec:growth}
The PAHs observed in disks are generally not believed to have been formed in situ after the collapse and main infall phases. Formation and growth of PAHs requires a high temperature ($\sim1000$ K), density and acetylene abundance \citep{cherchneff92a,frenklach89a} and, especially for the smallest PAHs, a low UV radiation field to prevent rapid photodissociation. If these conditions exist at all in a disk, it is only in a thin slice (less than $0.1$ AU) right behind the inner rim, and it is unlikely that this affects the PAH population at larger radii. Another possible method of late-stage PAH formation is in accretion shocks \citep{desch02a}, but too little is known about these events to include them in the model. Hence, no in situ formation or growth of PAHs is assumed to take place.

PAH destruction is governed by the loss of carbon fragments upon absorption of one or more UV photons. C--C bonds are a few eV stronger than C--H bonds (e.g. \tb{disspar}), so carbon is lost only from completely dehydrogenated PAHs. In order to get an accurate destruction rate, multi-photon events have to be taken into account, as first recognized by \citet{guhathakurta89a} and \citet{siebenmorgen92a}. Our model uses a procedure based on \citet{habart04a}. A PAH ensemble residing in a given radiation field assumes a statistical distribution over a range of internal energies. This distribution is represented by $P(E_\el{int})$, which is normalized so that $P(E_\el{int})\el{d}E_\el{int}$ is the probability to find the PAH in the energy interval $E_\el{int}$..$E_\el{int}+\el{d}E_\el{int}$. At every $E_\el{int}$, there is competition between cooling and dissociation with loss of a carbon fragment. The probability for dissociation depends on the ratio between $k_\el{C}$ (the sum of the rate constants for all carbon loss channels in \eq{kdiss}) and $k_\el{IR}$ (the instantaneous IR emission rate, comparable to \eq{kradi}):
\begin{equation}
\label{eq:etades}
\eta_\el{des}(E_\el{int}) = \frac{k_\el{C}(E_\el{int})}{k_\el{C}(E_\el{int}) + k_\el{IR}(E_\el{int})}\,.
\end{equation}
$k_\el{IR}$ is a very flat function in the relevant energy regime, while $k_\el{C}$ is very steep, so $\eta_\el{des}$ is approximately a step function.

The probability, $p_\el{des}$, that a PAH in a certain radiation field is destroyed is then found by integrating $P(E_\el{int})\eta_\el{des}(E_\el{int})$ over all energies:
\begin{equation}
\label{eq:pdes}
p_\el{des} = \int_{0}^{\infty} P(E_\el{int})\eta_\el{des}(E_\el{int})\el{d}E_\el{int}\,,
\end{equation}
with $P(E_\el{int})$ calculated according to \citet{guhathakurta89a}. This is equal to the formula in \citet{habart04a} if $\eta_\el{des}$ is replaced by a step function and the internal energy is converted to a temperature using the PAH's heat capacity \citep[e.g.][]{draine01a}. Destruction is assumed to take place if $p_\el{des}$ exceeds a value of $10^{-8}$.

\begin{figure}
\resizebox{\hsize}{!}{\includegraphics{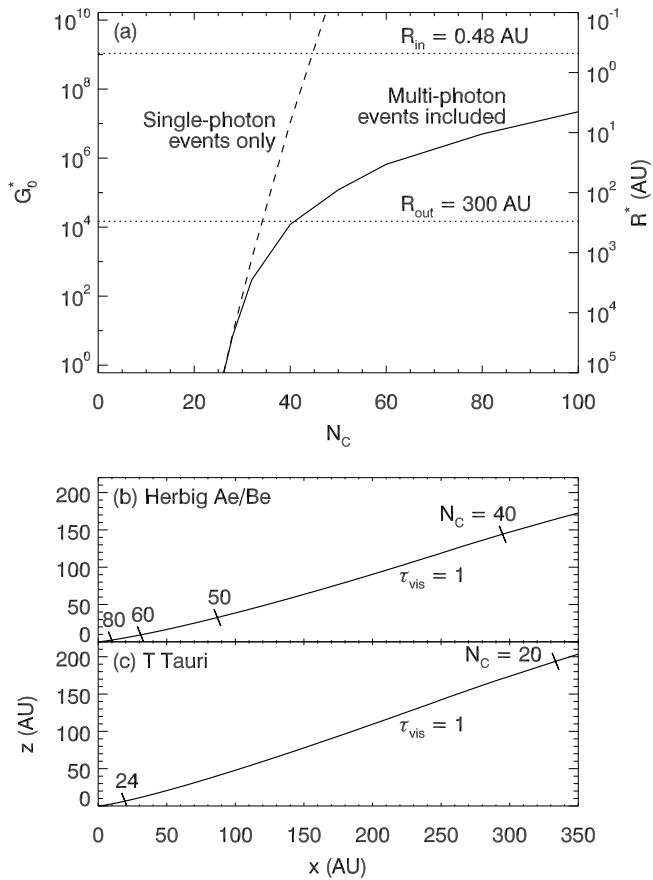}}
\caption{Photodestruction of PAHs. (a) The solid line gives $G_0^*$, the radiation intensity for which the destruction timescale is shorter than the disk lifetime of $3$ Myr, when multi-photon events are included. The dashed line gives $G_0^*$ when only single-photon destruction is allowed. The right vertical axis shows the corresponding ``destruction radius'' along the $\tau_\el{vis}=1$ surface in our model disk around a Herbig Ae/Be star (Section \ref{subsec:diskenv}); PAHs are destroyed inwards of this radius. The disk's inner and outer radius are indicated by the dotted lines. (b,c) The $\tau_\el{vis}=1$ surface in a vertical cut through our model Herbig Ae/Be and T Tauri disks. The tick marks denote the radius inwards of which PAHs of a given size are destroyed within the disk lifetime; for the Herbig Ae/Be model, this corresponds to the right vertical axis of panel (a). PAHs of equal size can survive much closer to a T Tauri star than to a Herbig Ae/Be star.}
\label{fig:photdes}
\end{figure}

We define a critical radiation intensity, $G_0^*$, which is the intensity required to cause photodestruction of a given PAH within a typical disk lifetime of $3$ Myr, i.e., the intensity required to get $\tau_\el{diss,C}=1/\Gamma_\el{diss,C}=3$ Myr. The $G_0^*$ for both single-photon and multi-photon destruction is plotted in \fig{photdes}. Knowledge of the radiation field at every point of the disk (\fig{c50radfield}) is required to determine where PAHs of a certain size are destroyed. An approximate approach is to trace the radiation field along the $\tau_\el{vis}=1$ surface, where the intensity will be shown to decrease almost as a power law (Section \ref{subsec:diskenv}). In our Herbig Ae/Be model (Section \ref{subsec:template}), PAHs of $50$ carbon atoms ($G_0^*=\scim{1.2}{5}$) are destroyed out to $100$ AU on the $\tau_\el{vis}=1$ surface. The destruction radius is larger for smaller PAHs and vice versa; e.g., PAHs with $\nc=100$ are only destroyed in the inner $5$ AU. PAHs with less than $24$ carbon atoms are not taken into account at all, because they are already destroyed when $G_0'\approx1$. Although there are regions inside the disk where the UV intensity is lower, the PAHs in such regions will not contribute significantly to the emission spectrum. For a T Tauri star, the radiation field in the disk is much weaker, so the destruction radius is smaller. In our model T Tauri disk, $50$-C PAHs can survive everywhere but in the disk's inner $0.01$ AU, while $100$-C PAHs can survive even there (Section \ref{subsec:ttauri}).

\subsection{Other chemical processes}
\label{subsec:otherchem}
No reactions between PAHs and species other than H and H$_2$ are included in our model. Although the second-order rate coefficients for the addition of, e.g., atomic nitrogen and oxygen are comparable to that for atomic hydrogen \citep[LPSB01;][]{snow98a}, the abundances of these heavier elements are not high enough to affect the chemical equilibrium. Formation of dimers and clusters \citep[and references therein]{rapacioli05b} and trapping of PAHs onto grains and ices \citep{gudipati03a} are also left out. The midplane of the disk, where densities are high enough and temperatures low enough for these processes to play a role, does not contribute significantly to the IR emission spectrum.


\section{Disk model}
\label{sec:diskmodel}


\subsection{Computational code}
\label{subsec:code}
The Monte Carlo radiative transfer code RADMC \citep{dullemond04a} is used in combination with the more general code RADICAL \citep{dullemond00a} to produce the IR spectra from PAHs in circumstellar disks. Using an axisymmetric density structure, but following photons in all three dimensions, RADMC determines the dust temperature and radiation field at every point of the disk. RADICAL then calculates a spectrum from all or part of the disk, or an image at any given wavelength.

The calculations in RADMC and RADICAL are based on the optical properties of a collection of carbon and silicate dust grains. Recently, PAHs were added as another type of grain to model the emission from the Herbig Ae star VV Ser and the surrounding nebulosity \citep{pontoppidan06a}, from a sample of Herbig Ae/Be and T Tauri stars \citep{geers06a}, and to study the effects of dust sedimentation (Dullemond et al. in prep.). The PAHs are excited in a quantized fashion by UV and, to a lesser degree, visible photons \citep{li02a}, and cool in a classical way according to the ``continuous cooling'' approximation \citep{guhathakurta89a}, which was found by \citet{draine01a} to be accurate even for small PAHs. A detailed description of the PAH emission module is given in \citet{pontoppidan06a}, whereas tests against other codes are described in \citet{geers06a}.

In the models used by \citet{pontoppidan06a}, \citet{geers06a} and Dullemond et al. (in prep.), no PAH chemistry was included. PAHs of a given size existed in the same charge/hydrogenation state everywhere, or in a fixed ratio between a limited number of states (e.g. $50\%$ neutral, $50\%$ ionized). In the current model, the chemistry is included in the following way. First, a single charge/hydrogenation state for a given $\nc$ is included at a given abundance in the radiative transfer procedure, to calculate the disk structure and radiation field at every point. Using these physical parameters, the equilibrium distribution of the PAHs over all possible charge/hydrogenation states is then determined. After an optional second iteration of the radiative transfer to take into account the heating of thermal grains by emission from the PAHs, the spectrum or image is calculated.

Some additional chemistry is added to the model in order to determine the electron and atomic and molecular hydrogen densities, which are needed to calculate the chemical equilibrium of the PAHs. The electron abundance, $x_\el{e}=n_\el{e}/n_\el{H}$, is set equal to the C$^+$ abundance, based on a simple equilibrium between the photoionization of neutral C and the recombination of C$^+$ \citep{bergin06a_ppv,bergin03a,leteuff00a}. All hydrogen is in atomic form at the edges of the disk and is converted to molecular form inside the disk as the amount of dissociating photons decreases due to self-shielding and shielding by dust \citep{vanzadelhoff03a,draine96a}. The H$_2$ formation rate is taken from \citet{black87a}. For the outer parts of the disk, which receive little radiation from the star, an interstellar radiation field with $G_0=1$ is included.

The RADMC and RADICAL codes only treat isotropic scattering of photons. We verified with a different radiative transfer code \citep{vanzadelhoff03a} that no significant changes occur in the results when using a more realistic anisotropic scattering function.

Different PAHs (i.e., different sizes) can be included at the same time, and for each one, the equilibrium distribution over all charge/hydrogenation states is calculated. The abundance of each PAH with a given $\nc$ is equal throughout the disk, except in those regions where $G_0'>G_0^*(\nc)$; there, the abundance is set to zero. As discussed in Section \ref{subsec:growth}, mixing processes are ignored.


\subsection{Template disk with PAHs}
\label{subsec:template}
For most of the calculations, a template Herbig Ae/Be star+disk model is used with the following parameters. The star has radius $2.79R_{\sun}$, mass $2.91M_{\sun}$ and effective temperature $10^4$ K, and its spectrum is described by a Kurucz model. No UV excess due to accretion or other processes is present. The mass of the disk is $0.01M_{\sun}$, with inner and outer radii of $0.48$ and $300$ AU. The inner radius corresponds to a dust evaporation temperature of $1700$ K. The disk is in vertical hydrostatic equilibrium, with a flaring shape and a slightly puffed-up inner rim \citep{dullemond04a}. The dust temperature is calculated explicitly, whereas the gas temperature is put to a constant value of $300$ K everywhere, appropriate for the upper layers from which most of the PAH emission originates. The results are not sensitive to the exact value of the gas temperature (see also Section \ref{subsec:sens}).

The same parameters are used for a template T Tauri star+disk model, except that the star has mass $0.58M_{\sun}$ and effective temperature $4000$ K. The dust evaporation temperature is kept at $1700$ K, so the disk's inner radius moves inwards to $0.077$ AU.

$\ch{50}{18}{}$ is present as a prototypical PAH, at a high abundance of \scit{1.6}{-6} PAH molecules per hydrogen nucleus to maximize the effects of changes in the model parameters. This abundance corresponds to $10\%$ of the total dust mass and to $36\%$ of the total amount of carbon in dust being locked up in this PAH, assuming an abundance of carbon in dust of $2.22\times10^{-4}$ with respect to hydrogen \citep{habart04a}. The model was also run for PAHs of $24$ and $96$ carbon atoms.


\section{Results}
\label{sec:results}

This work focuses on the chemistry of the PAHs in a circumstellar disk and on its effects on the mid-IR emission, as well as on the differences between disks around Herbig Ae/Be and T Tauri stars. \citet{geers06a} analyzed the effects of changing various disk parameters, such as the PAH abundance and the disk geometry; settling of PAHs and dust will be analyzed by Dullemond et al. (in prep.).


\subsection{PAH chemistry}
\label{subsec:pahdis}
Due to the large variations in density and UV intensity throughout the disk, the PAHs are present in a large number of charge/hydrogenation states (\tb{stst}). When a disk containing only $\ch{50}{y}{}$ ($\nho=18$) is in steady state, our model predicts that $56\%$ of the observed PAH emission between $2.5$ and $13.5$ \micron{} originates from $\ch{50}{18}{}$, with another $9\%$ from $\ch{50}{18}{+}$. PAHs missing one hydrogen atom also contribute to the emission: $22\%$ comes from $\ch{50}{17}{}$ and $12\%$ from $\ch{50}{17}{+}$.

\begin{figure}
\resizebox{\hsize}{!}{\includegraphics{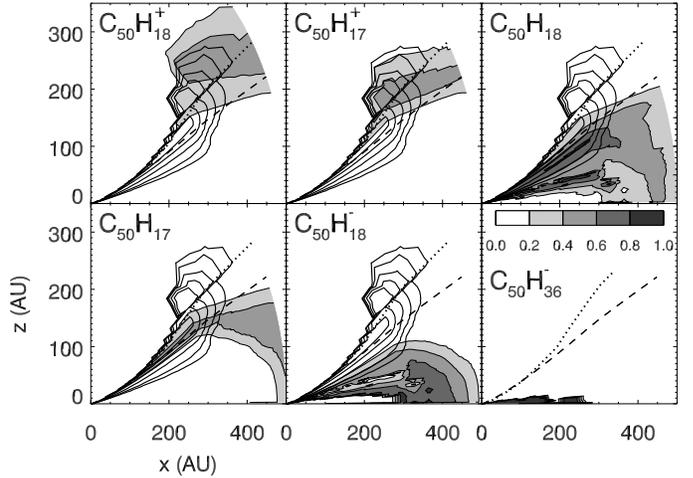}}
\caption{Steady-state distribution of the most important charge/hydrogenation states of $\ch{50}{y}{}$ ($\nho=18$) in a disk around a Herbig Ae/Be star ($10^4$ K). Each panel shows a cut through the disk, with the equator on the $x$ axis and the pole on the $z$ axis. The gray scale denotes the fraction at which a state is present compared to all possible states. The dashed line denotes the $\tau_\el{vis}=1$ surface. The dotted ``emission line'' connects the points where the PAH emission is strongest for a given distance $R=\sqrt{x^2+z^2}$ from the star. The thick black contour lines denote the region responsible for most of the PAH emission; from the outside inward, the contours contain $95$, $90$, $76$, $52$ and $28\%$ of the total emitted power between $2.5$ and $13.5$ \micron. They are omitted from the last frame for clarity.}
\label{fig:c50dis}
\end{figure}

A strong contribution to the observed emission does not imply a high abundance throughout the disk, because the contribution of each state to the spectrum also depends on its spatial distribution. \figg{c50dis} shows a cut through the disk and indicates the abundance of the six most important states with respect to the total $\ch{50}{y}{}$ population. Near the surface, where the radiation field is strong, the PAHs are ionized and some of them have lost a hydrogen atom. Going to lower altitudes, the PAHs first become neutral and then negatively ionized. Still lower, the increasing density and optical depth lead to further hydrogenation, resulting in a high abundance of the completely hydrogenated anion, $\ch{50}{36}{-}$, around the midplane.

Since normally hydrogenated and positively ionized states occur in more optical/UV-intense regions than do completely hydrogenated and negatively ionized states, the former will emit more strongly. For example, the state responsible for more than half of the emission, $\ch{50}{18}{}$, constitutes only $21\%$ of all PAHs in the entire disk (\tb{stst}). About half of the PAHs ($45\%$) are expected to be in the form of $\ch{50}{36}{-}$. This state, which is assumed not to emit at all (Section \ref{subsec:cross}), dominates the high-density regions close to the midplane. The normally hydrogenated anion, $\ch{50}{18}{-}$, accounts for $28\%$ of all PAHs, and the remaining PAHs are mostly present as anions with $18<\nh<36$. States like $\ch{50}{17}{-}$ and $\ch{50}{36}{+}$ are entirely absent: the same environmental parameters that favour dehydrogenation (strong radiation, low density), also favour ionization.

Species with $\nh<17$ are also predicted to be absent. They have to be formed from $\ch{50}{17}{Z}$, but wherever these states exist, the ratio between the UV intensity and the hydrogen density favours hydrogen addition. In those regions where the ratio is favourable to photodissociation with H loss, the radiation field is also strong enough to destroy the carbon skeleton. The abundance of $\ch{50}{17}{Z}$ is boosted by H$_2$ loss from $\ch{50}{19}{Z}$, which is a much faster process than H loss from $\ch{50}{18}{Z}$. Moving from the surface to the midplane, the increasing $n_\el{H}/G_0'$ ratio allows all hydrogenation states from $\nho+1$ to $2\nho$ to exist. However, since all of them except $\nh=19$ are negatively charged, they are assumed not to emit.

\subsection{PAH emission}
\label{subsec:diskenv}
\begin{figure}
\resizebox{\hsize}{!}{\includegraphics{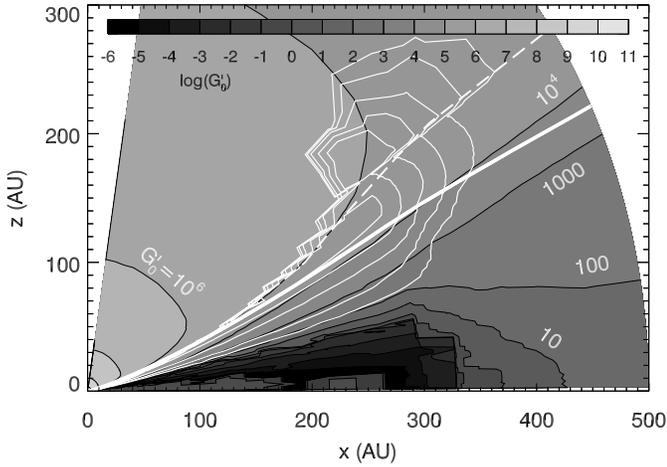}}
\caption{The radiation field (characterized by $G_0'$; \eq{g_0}) for a flaring disk around a Herbig Ae/Be star ($10^4$ K), containing PAHs of $50$ carbon atoms. The dashed line is the ``emission line'' from \fig{c50dis}. The white contour lines are the same as the black contour lines in \fig{c50dis}.}
\label{fig:c50radfield}
\end{figure}

\begin{figure}
\resizebox{\hsize}{!}{\includegraphics{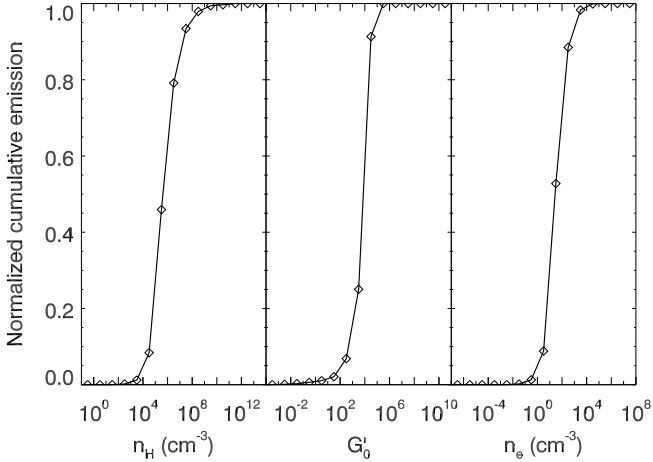}}
\caption{Normalized cumulative PAH emission for the Herbig Ae/Be model as function of total hydrogen density ($n_\el{H}$, left panel), radiation field ($G_0'$, middle panel) and electron density ($n_\el{e}$, right panel).}
\label{fig:c50emisvs}
\end{figure}

The integrated UV intensity between $6$ and $13.6$ eV, characterized by $G_0'$ (\eq{g_0}), varies by many orders of magnitude from the disk's surface to the midplane. The radiation field for the model star+disk system is shown in \fig{c50radfield}, with the region responsible for most of the $2.5$--$13.5$ \micron{} PAH emission indicated by contour lines. Part of this region is truncated sharply due to photodestruction of the PAHs, marking the $G_0'=G_0^*$ line. An analysis of the disk structure shows that some $95\%$ of the PAH emission originates from a region with $10^1<G_0'<10^5$, $10^3<n_\el{H}\ (\el{cm}^{-3})<10^9$, and $10^{-1}<n_\el{e}\ (\el{cm}^{-3})<10^4$ (see also \fig{c50emisvs}).

When the radiation field is traced along the $\tau_\el{vis}=1$ surface, and the disk's inner $0.01$ AU is disregarded, the integrated intensity decreases almost as a power law. Specifically, for any point on the $\tau_\el{vis}=1$ surface a distance $R_\tau$ (in AU) away from the star,
\begin{equation}
\label{eq:g0tau}
G_0' \approx 3.0\times10^8 R_\tau^{-1.74}\,.
\end{equation}
The power-law exponent differs from the value of $-2$ expected based on geometrical considerations because of the curvature in the $\tau_\el{vis}=1$ surface. This relationship allows one to predict to what radius $R_\tau^*$ PAHs of a certain size are destroyed, as was done in Section \ref{subsec:growth} and \fig{photdes}. It should be noted, however, that this only applies to the $\tau_\el{vis}=1$ surface. PAHs can survive and contribute to the emission from $R<R_\tau^*$ when they are at lower altitudes (e.g. Figs.~\ref{fig:c50radfield} and \ref{fig:spex}).

The abundances of the charge/hydrogenation states in the $95\%$ emission region are different than those in the entire disk. Although they agree more closely to the emissivity contributions, a perfect correspondence is not achieved. For instance, only $1.0\%$ of the PAHs in this region are in the form of $\ch{50}{18}{+}$, but they account for $8.0\%$ of the emission (\tb{stst}). This is again due to the positive ions occurring in a more intense radiation field than the neutrals and negative ions.

In a vertical cut through the disk, a point exists for every distance $R_\el{em}$ from the star where the PAH emission is strongest. These points form the ``emission lines'' in Figs.~\ref{fig:c50dis} and \ref{fig:c50radfield}. The conditions along this line determine the charge and hydrogenation of the PAHs responsible for most of the observed emission. Its altitude depends mostly on two competing factors: the intensity of the UV field and the PAH density. A strong UV field leads to stronger emission, but a strong UV field can only exist in regions of low density, where the total emission is weaker.

\begin{figure}
\resizebox{\hsize}{!}{\includegraphics{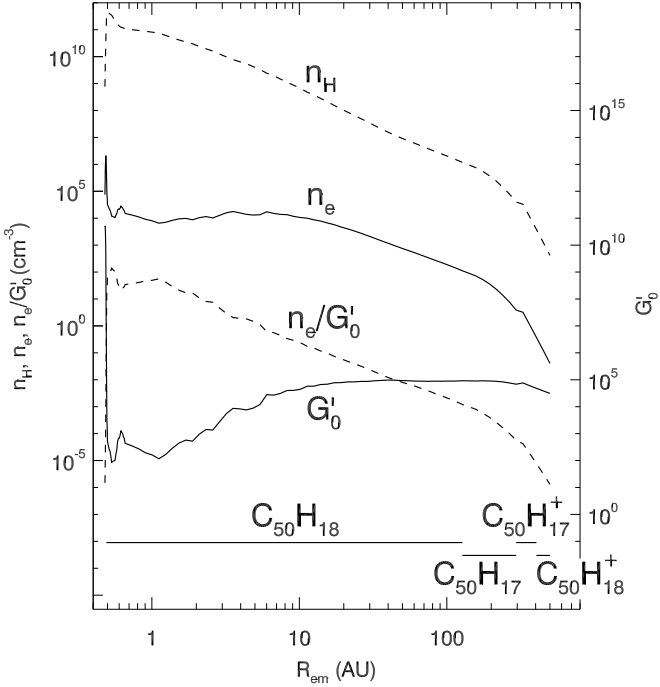}}
\caption{Total hydrogen density ($n_\el{H}$), electron density ($n_\el{e}$), radiation field ($G_0'$) and ratio between electron density and radiation field ($n_\el{e}/G_0'$) along the emission line defined in \fig{c50dis} and Section \ref{subsec:diskenv}. The bars at the bottom indicate the main emitter at each distance along the emission line. The parameters are those of the template Herbig Ae/Be model.}
\label{fig:emline}
\end{figure}

The variation of $n_\el{H}$, $n_\el{e}$ and $G_0'$ along the emission line is plotted in \fig{emline}. In the disk's inner $100$ AU (disregarding the actual inner rim), photodestruction of PAHs causes the emission line to lie below the $\tau_\el{vis}=1$ surface, where $n_\el{H}$ and $n_\el{e}$ are relatively high ($10^{11}$ and $10^4$ \pcc, respectively) and $G_0'$ is relatively low ($10^2$--$10^3$). The resulting $n_\el{e}/G_0'$ and $n_\el{H}/G_0'$ ratios favour the neutral normally hydrogenated species, $\ch{50}{18}{}$. Going to larger radii, the emission line gradually moves up and crosses the $\tau_\el{vis}=1$ surface at around $100$ AU. As a consequence, $n_\el{H}$ and $n_\el{e}$ decrease while $G_0'$ increases. Because the electron abundance with respect to hydrogen also decreases, the PAHs first lose a hydrogen atom to become $\ch{50}{17}{}$. Further out, ionization takes place to produce $\ch{50}{17}{+}$. At still larger radii, the radiation field starts to lose intensity and $\ch{50}{18}{+}$ becomes the dominant species. The deviation from the $R^{-1.74}$ power law for the radiation field (\eq{g0tau}) is due to the emission line not following the $\tau_\el{vis}=1$ surface.

It should be noted that ionized and/or dehydrogenated species do exist (and emit) in the disk's inner part. However, for any given distance to the star, they only become the most abundant species at around $100$ AU.

\subsection{Other PAHs}
\label{subsec:otherpahs}
If $\ch{96}{24}{}$ is put into the disk instead of $\ch{50}{18}{}$, the abundance of the dehydrogenated and normally hydrogenated states essentially goes to zero. This is due to the hydrogen addition rates being larger and the hydrogen dissociation rates being smaller for larger PAHs. Most of the $2.5$--$13.5$ \micron{} emission in the $\nc=96$ case comes from $\ch{96}{48}{}$ ($44\%$), $\ch{96}{48}{+}$ ($42\%$) and $\ch{96}{48}{2+}$ ($14\%$), while the anion, $\ch{96}{48}{-}$, is the most abundant overall (\tb{stst}).

\begin{figure}
\resizebox{\hsize}{!}{\includegraphics{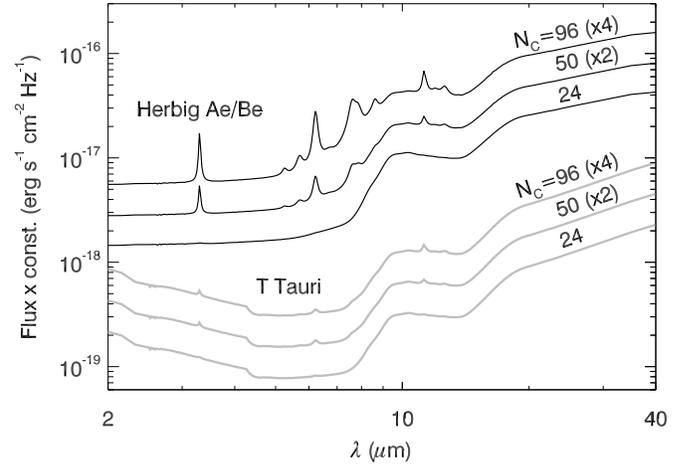}}
\caption{Model spectra (flux at $1$ pc) for disks around a Herbig Ae/Be (black) and a T Tauri star (gray) containing either $\ch{24}{12}{}$, $\ch{50}{18}{}$ (shifted by a factor of $2$) or $\ch{96}{24}{}$ (shifted by a factor of $4$), with the charge and hydrogenation balance calculated by the chemistry model.}
\label{fig:allspec}
\end{figure}

If only $\ch{24}{12}{}$ is put in, the PAH emission becomes very weak. The critical radiation intensity for this PAH is less than unity (\fig{photdes}), so it will only survive in strongly shielded areas. There, $66\%$ of the PAHs are present as $\ch{24}{12}{-}$, $28\%$ as $\ch{24}{12}{}$ and $4\%$ as $\ch{24}{24}{-}$. Only the neutral species contributes to the emission; however, since this PAH only emits from regions where $G_0'\la1$, no PAH features are visible in the calculated spectrum, despite the high abundance used in our model. This is exemplified in \fig{allspec}, where the calculated spectra for the model star+disk system are compared for the three PAH sizes. A disk around a Herbig Ae/Be star containing PAHs of $50$ or $96$ carbon atoms shows strong PAH features, but the $24$-C spectrum contains only thermal dust emission.

The goal of \fig{allspec} is to show the differences arising from the photochemical modelling of the three PAH sizes in model Herbig Ae/Be and T Tauri disks, rather than to provide realistic spectra from such objects. In order to fit our model results to observations, one would need to include a range of PAH sizes in one model \citep{li03b} and execute a larger parameter study \citep{geers06a,habart04a} than was done in this work.


\subsection{Spatial extent of the PAH emission}
\label{subsec:spex}
\begin{figure}
\resizebox{\hsize}{!}{\includegraphics{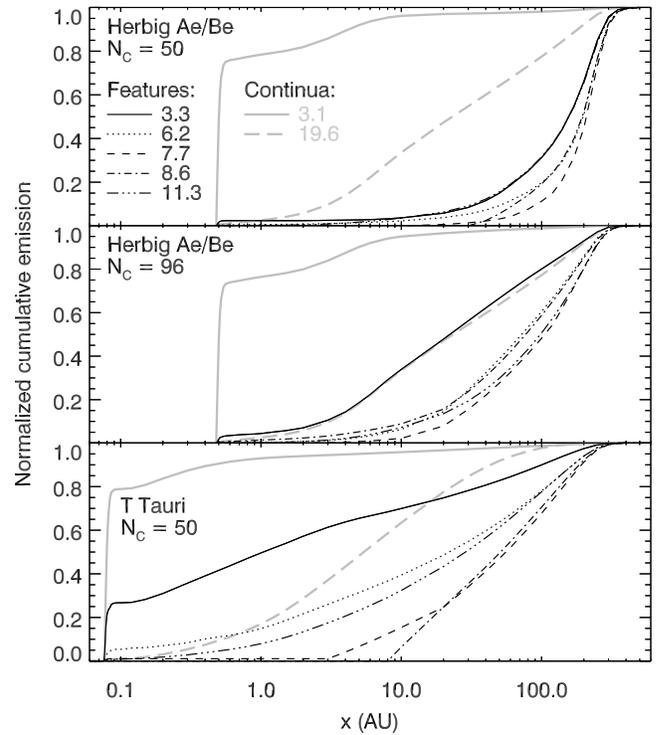}}
\caption{Normalized cumulative integrated intensity of the five main PAH features (black) and the continua at $3.1$ and $19.6$ \micron{} (gray) for (top) $\ch{50}{18}{}$ around a Herbig Ae/Be star, (middle) $\ch{96}{24}{}$ around a Herbig Ae/Be star and (bottom) $\ch{50}{18}{}$ around a T Tauri star. In each case, the charge and hydrogenation balance were calculated by the chemistry model.}
\label{fig:spex}
\end{figure}

The top panel in \figg{spex} presents the cumulative intensity of the five main PAH features and the continua at $3.1$ and $19.6$ \micron{} as a function of radius for $\nc=50$. In accordance with Figs.~\ref{fig:c50dis} and \ref{fig:c50radfield}, more than $95\%$ of the power radiated in the features originates from outside the inner $10$ AU, and some $80\%$ from outside $100$ AU. This is largely due to these PAHs being destroyed closer to the star. The $3.3$ and $11.3$ \micron{} features are somewhat less extended than the other three features. The continua are much more confined than the features, especially at $3.1$ \micron{}, where $75\%$ comes from the disk's inner rim.

The PAH emission from a disk with $\nc=96$ (\fig{spex}, middle panel) is less extended than that from the same disk with $\nc=50$, because larger PAHs can survive at smaller radii. About $80\%$ of the integrated intensity of the $3.3$ \micron{} feature originates from within $100$ AU, and so does about $60\%$ of the other, less energetic features. The continuum at $20$ \micron{} has the same spatial behaviour as the $3.3$ \micron{} PAH feature, while the continuum at $3.1$ \micron{} is very much confined towards the center. These results are in good agreement with the spatial extent modelled by \citet{habart04a}. The spatial extent of the continuum emission from the $\ch{96}{24}{}$ disk is identical to that from the $\ch{50}{18}{}$ disk, so the thermal dust emission appears to be unaffected by the details of the PAHs and their chemistry.

Our model also agrees well with a number of spatially resolved observations \citep{habart06a,geers05a,vanboekel04a}, where the PAH features were consistently found to be more extended than the adjacent continuum. Furthermore, PAH emission is typically observed on scales of tens of AUs, which cannot be well explained by our model if only PAHs of $50$ carbon atoms are present. Hence, the observed emission is probably due to PAHs of at least about $100$ carbon atoms.


\subsection{Sensitivity analysis}
\label{subsec:sens}
The rates for most of the chemical reactions discussed in Section \ref{sec:pahmodel} are not yet well known. In order to gauge the importance of having an accurate rate for a given reaction, the equilibrium distributions were calculated for rates increased or decreased by a factor of $100$ from their normal model values. These results are also presented in \tb{stst}. Modifying the hydrogen dissociation and addition rates leads to no significant changes, so treating photodissociation with loss of hydrogen in a purely single-photon fashion likely does not introduce large errors. The fact that most of the PAH emission comes from regions with a relatively weak UV field ($G_0'<10^5$), where multi-photon events play only a minor role, further justifies the single-photon dehydrogenation treatment.

The ionization and electron attachment rates are more important to know accurately. For instance, the contribution from the cations to the spectrum decreases by a factor of a few when taking an ionization rate that is $0.01$ times the normal model rate. The shift away from positively charged species also affects hydrogenation (addition of hydrogen is faster to cations than to neutrals), resulting, e.g., in a smaller emissivity contribution from $\ch{50}{18}{}$ and $\ch{50}{18}{+}$ with respect to $\ch{50}{17}{}$ and $\ch{50}{17}{+}$. Further laboratory work on ionization and electron attachment rates, especially for larger PAHs, will help to better constrain this part of the model.

As shown by e.g. \citet{kamp04a} and \citet{jonkheid04a}, the gas temperature is not constant throughout the disk. The recombination rates between electrons and cations decrease for higher temperatures, whereas the attachment rates of electrons to neutrals are almost independent of temperature (Section \ref{subsec:elatt}). Taking a higher temperature would result in a slightly larger cation abundance. However, even for an extreme temperature of $5000$ K, the recombination rates decrease by only a factor of $\sim4$, and the effects will be smaller than what is depicted in \tb{stst}. Hence, we believe it is justified to take a constant temperature of $300$ K throughout the disk.


\subsection{T Tauri stars}
\label{subsec:ttauri}


The results discussed so far are for a star with an effective temperature of $10^4$ K, appropriate for a Herbig Ae/Be type. Colder stars, like T Tauri types, are less efficient in inducing IR emission in PAHs, as shown with both observations and models by \citet{geers06a}, unless they have excess UV over the stellar atmosphere. Hence, when the model parameters are otherwise unchanged, the observed flux from the star+disk system becomes lower and the absolute PAH features become weaker (\fig{allspec}).

It was shown in Section \ref{subsec:diskenv} that the integrated intensity of the UV field along the $\tau_\el{vis}=1$ surface of a disk around a Herbig Ae/Be star decreases approximately as a power law. This is also true for the T Tauri case, if the disk's inner $0.01$ AU are again disregarded. The exponent is slightly larger:
\begin{equation}
\label{eq:g0tautt}
G_0' \approx 12 R_\tau^{-1.93}\,,
\end{equation}
with $R$ in AU. The UV field peaks at $G_0'=\scim{6}{5}$ at the inner rim and drops by two orders of magnitude within $0.01$ AU, so a PAH of $50$ carbon atoms ($G_0^*=\scim{1.2}{5}$) can survive almost everywhere (see also \fig{photdes}).

The lower destruction radii also lead to the PAH emission being more concentrated towards the inner disk (\fig{spex}, bottom panel). The $3.3$ \micron{} feature is particularly confined, with $70\%$ originating from the inner $10$ AU and $25\%$ from the inner rim. As was the case for the Herbig Ae/Be disks (Section \ref{subsec:spex}), the inner rim contributes strongly to the continuum emission at $3.1$ \micron{}. The $19.6$ \micron{} continuum also behaves very similarly to that from the Herbig Ae/Be disks, except that it is stretched inwards because of the smaller $R_\el{in}$. Thus, the spatial extent of the thermal dust emission seems to be largely unaffected by the temperature of the central star.

Cations are practically absent in the model T Tauri disk with any kind of PAH, due to the weaker radiation field. For $\nc=24$ and $50$, almost all of the PAH emission originates from the normally hydrogenated neutral species, $\ch{24}{12}{}$ or $\ch{50}{18}{}$. Less than $0.1\%$ originates from other neutral states, primarily those missing one hydrogen atom ($\ch{24}{11}{}$, $\ch{50}{17}{}$) or having twice the normal number of hydrogen atoms ($\ch{24}{24}{}$, $\ch{50}{36}{}$). If a disk around a T Tauri star contains only $96$-C PAHs, all of the emission is due to $\ch{96}{48}{}$. The absence of cations could help explain the weak $7.7$ and $8.6$ \micron{} features in observed spectra \citep{geers06a}, because these two features are weaker in neutral PAHs.

Anions are abundant for all three PAH sizes, accounting for about half of the entire PAH population. However, they are again assumed not to contribute to the emission.


\subsection{Comparison with observations}
\label{subsec:compobs}

\begin{figure}
\resizebox{\hsize}{!}{\includegraphics{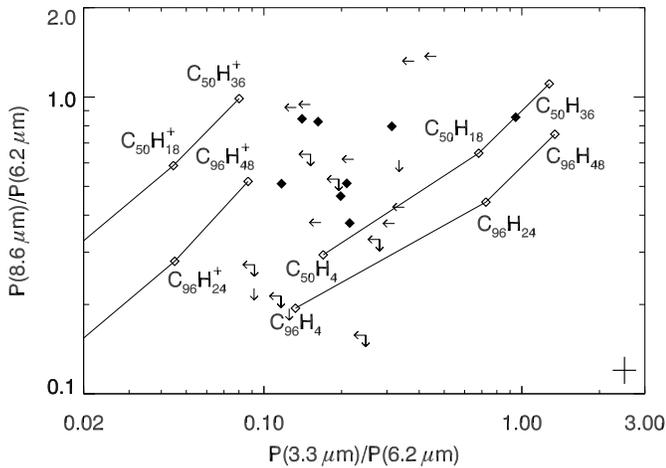}}
\caption{The ratio of the fluxes of the $8.6$ and $6.2$ \micron{} bands against the ratio of the fluxes of the $3.3$ and $6.2$ \micron{} bands. Filled diamonds and arrows are detections and upper limits from \citet{acke04a}. Open diamonds and lines are predictions from the template Herbig Ae/Be model. The cross in the lower right shows the typical error bars for both the observations and the model results.}
\label{fig:8663}
\end{figure}

\citet{acke04a} performed a comprehensive analysis of the PAH features in a large sample of Herbig Ae/Be stars observed with ISO, measuring line fluxes and comparing them to each other. \figg{8663} recreates their Fig.~9, plotting the ratio of the integrated fluxes in the $8.6$ and $6.2$ \micron{} bands against the ratio of the integrated fluxes in the $3.3$ and $6.2$ \micron{} bands. In order to gauge the plausibility of the numerous charge/hydrogenation states a PAH can in principle attain, the same ratios are also plotted for a sample of models containing one PAH in one specific state only.

The $3.3/6.2$ ratio from the model is very sensitive to the charge of the PAH and increases by an order of magnitude when going from ionized to neutral species. The $3.3$ and $8.6$ \micron{} features are due to C--H vibrational modes, while the $6.2$ \micron{} feature is due to a C--C mode (Section \ref{subsec:cross}), so both ratios in \fig{8663} increase with $\nh$. The observations fall mostly in between the model points for neutral and ionized PAHs, so both charge states appear to contribute to the observed emission. This strengthens the model results presented in this work, although the observed emissivity contribution from neutral species seems to be somewhat less than the predicted $~50\%$. Most of the observations agree with the model prediction that the emission originates from multiple hydrogenation states, with a lower limit of $\nh=\nho-1$.

\section{Conclusions}
\label{sec:con}
The chemistry of, and infrared (IR) emission from, polycyclic aromatic hydrocarbons (PAHs) in disks around Herbig Ae/Be and T Tauri stars were studied. An extensive PAH chemistry model has been created, based primarily on the models of \citet{lepage01a} and \citet{weingartner01a}, with absorption cross sections from \citet{draine06a}. This model includes reactions affecting the charge (ionization, electron recombination, electron attachment) and hydrogen coverage (photodissociation with hydrogen loss, hydrogen addition) of PAHs in an astronomical environment. Destruction of PAHs by UV radiation is also taken into account, including destruction by multi-photon absorption events. By coupling the chemistry model to an existing radiative transfer model, equilibrium charge/hydrogenation distributions throughout the disks were obtained. The main results are as follows:

\begin{itemize}
\item[$\bullet$] Very small PAHs ($24$ carbon atoms) are destroyed within a typical disk lifetime of $3$ Myr even in regions of low UV intensity ($G_0'\sim1$). No features are seen in the calculated spectrum for either a Herbig Ae/Be or a T Tauri disk, despite a high PAH abundance.

\item[$\bullet$] PAHs of intermediate size ($50$ carbon atoms) do produce clearly visible features, even though they are still photodestroyed out to about $100$ AU in the surface layers of a disk around a Herbig Ae/Be star. The model predicts that most of the emission arises from the surface layers and from large radii (more than $100$ AU). Neutral and positively ionized species, bearing the normal number of hydrogen atoms or one less, contribute in roughly equal amounts. Negatively charged species are also present, but are assumed not to contribute to the emission.

\item[$\bullet$] Going to still larger PAHs ($96$ carbon atoms), photodestruction becomes a slower process and the PAHs can survive down to $5$ AU from a Herbig Ae/Be star. The slower photodissociation rates also mean that these PAHs are fully hydrogenated everywhere in the disk. Neutral and ionized species still contribute in comparable amounts to the emission, with some $15\%$ originating from doubly ionized PAHs.

\item[$\bullet$] The PAH emission is predicted to be extended on a scale similar to the size of the disk, with the features at longer wavelengths contributing more in the outer parts and the features at shorter wavelengths contributing more in the inner parts. For similar wavelengths, the continuum emission is less extended than the PAH emission.

\item[$\bullet$] Disks around T Tauri stars show weaker PAH features than do disks around Herbig Ae/Be stars because of the weaker radiation from T Tauri stars, assuming they have no excess UV over the stellar atmosphere. The PAH emission from T Tauri disks is considerably more confined towards the center than that from disks around Herbig Ae/Be stars, because PAHs can survive much closer to the star. For instance, a $50$-C PAH survives everywhere but in the disk's innermost $0.01$ AU. Furthermore, the radiation field is no longer strong enough to ionize the PAHs, and all the PAH emission originates from neutral species for all three PAH sizes. This could help explain the weak $7.7$ and $8.6$ \micron{} features in observed spectra. About half of all PAHs in a T Tauri disk are predicted to be negatively ionized.

\item[$\bullet$] Comparing the model results to spatially resolved observations \citep{habart06a,geers05a,vanboekel04a} for Herbig Ae/Be stars, it appears that PAHs of at least about $100$ carbon atoms are responsible for most of the emission. The emission from smaller species is predicted to be too extended. Other observations \citep{acke04a} support the conclusion that the emission is due to a mix of neutral and singly positively ionized species.
\end{itemize}


\begin{acknowledgements}
The authors are grateful to Xander Tielens and Steven Doty for stimulating discussions. Astrochemistry in Leiden is supported by an NWO Spinoza Grant and a NOVA grant, and by the European Research Training Network ``The Origin of Planetary Systems'' (PLANETS, contract number HPRN-CT-2002-00308). Support for KMP was provided by NASA through Hubble Fellowship grant \#01201.01 awarded by the Space Telescope Science Institute, which is operated by the Association of Universities for Research in Astronomy, Inc., for NASA, under contract NAS 5-26555.
\end{acknowledgements}


\bibliographystyle{aa}
\bibliography{pahchem}

\begin{longtable}{l l c ccccc c ccccc}
\caption{\label{tb:stst} Abundances (columns 3 to 7) and $2.5$--$13.5$ \micron{} emissivity contributions (columns 8 to 12) of the dominant charge/hydrogenation states for three PAHs in the model Herbig Ae/Be star+disk system.$^{\mathrm{a,b}}$}\\
\hline\hline
      &       & & \multicolumn{5}{c}{Abundance in entire disk (\%)$^{\mathrm{c}}$}                                        & & \multicolumn{5}{c}{Contribution to $2.5$--$13.5$ \micron{} emission (\%)$^{\mathrm{c}}$}                \\
\hline
$\nh$ & $\,Z$ & & N.R. &  $100\Gamma_\el{ea}$ & $0.01\Gamma_\el{ea}$ & $100\Gamma_\el{diss,H}$ & $0.01\Gamma_\el{diss,H}$ & & N.R. &  $100\Gamma_\el{ea}$ & $0.01\Gamma_\el{ea}$ & $100\Gamma_\el{diss,H}$ & $0.01\Gamma_\el{diss,H}$ \\
      &       & &      & $0.01\Gamma_\el{em}$ &  $100\Gamma_\el{em}$ & $0.01\Gamma_\el{add,H}$ &  $100\Gamma_\el{add,H}$  & &      & $0.01\Gamma_\el{em}$ &  $100\Gamma_\el{em}$ & $0.01\Gamma_\el{add,H}$ &  $100\Gamma_\el{add,H}$  \\
\hline
\endfirsthead
\caption{continued.}\\
\hline\hline
      &       & & \multicolumn{5}{c}{Abundance in entire disk (\%)$^{\mathrm{c}}$}                                        & & \multicolumn{5}{c}{Contribution to $2.5$--$13.5$ \micron{} emission (\%)$^{\mathrm{c}}$}                \\
\hline
$\nh$ & $\,Z$ & & N.R. &  $100\Gamma_\el{ea}$ & $0.01\Gamma_\el{ea}$ & $100\Gamma_\el{diss,H}$ & $0.01\Gamma_\el{diss,H}$ & & N.R. &  $100\Gamma_\el{ea}$ & $0.01\Gamma_\el{ea}$ & $100\Gamma_\el{diss,H}$ & $0.01\Gamma_\el{diss,H}$ \\
      &       & &      & $0.01\Gamma_\el{em}$ &  $100\Gamma_\el{em}$ & $0.01\Gamma_\el{add,H}$ &  $100\Gamma_\el{add,H}$  & &      & $0.01\Gamma_\el{em}$ &  $100\Gamma_\el{em}$ & $0.01\Gamma_\el{add,H}$ &  $100\Gamma_\el{add,H}$  \\
\hline
\endhead
\hline
\endfoot
\multicolumn{14}{c}{$\nc=24$, $\nho=12$, $-1\le{}Z\le+2$} \\
$12$           & $-1$           & & $6.6(+1)$ & $9.2(+1)$ & $3.5(+0)$ & $6.7(+1)$ & $6.5(+1)$ & & $-      $ & $-      $ & $-      $ & $-      $ & $-      $ \\
$12$           & $\phantom{+}0$ & & $2.8(+1)$ & $6.0(-1)$ & $8.9(+1)$ & $2.7(+1)$ & $2.8(+1)$ & & $1.0(+2)$ & $1.0(+2)$ & $1.0(+2)$ & $1.0(+2)$ & $1.0(+2)$ \\
$13$           & $-1$           & & $1.1(-1)$ & $1.8(-1)$ & $2.6(-4)$ & $1.1(-1)$ & $2.8(-1)$ & & $-      $ & $-      $ & $-      $ & $-      $ & $-      $ \\
$14$           & $-1$           & & $1.0(-1)$ & $2.4(-3)$ & $2.4(-4)$ & $1.1(-1)$ & $1.1(-1)$ & & $-      $ & $-      $ & $-      $ & $-      $ & $-      $ \\
$15$           & $-1$           & & $1.0(-1)$ & $2.6(-3)$ & $2.4(-4)$ & $1.1(-1)$ & $1.1(-1)$ & & $-      $ & $-      $ & $-      $ & $-      $ & $-      $ \\
$16$           & $-1$           & & $1.0(-1)$ & $2.9(-3)$ & $2.4(-4)$ & $1.0(-1)$ & $1.1(-1)$ & & $-      $ & $-      $ & $-      $ & $-      $ & $-      $ \\
$23$           & $\phantom{+}0$ & & $1.0(-1)$ & $6.4(-5)$ & $6.3(-3)$ & $1.0(-1)$ & $1.0(-1)$ & & $1.2(-3)$ & $-      $ & $2.7(-5)$ & $1.9(-3)$ & $1.2(-3)$ \\
$24$           & $-1$           & & $3.6(+0)$ & $7.4(+0)$ & $3.7(-1)$ & $3.7(+0)$ & $3.6(+0)$ & & $-      $ & $-      $ & $-      $ & $-      $ & $-      $ \\
$24$           & $\phantom{+}0$ & & $7.3(-1)$ & $5.1(-2)$ & $7.1(+0)$ & $7.5(-1)$ & $7.3(-1)$ & & $3.6(-2)$ & $3.6(-1)$ & $5.1(-2)$ & $5.5(-2)$ & $3.6(-2)$ \\
\hline
\multicolumn{14}{c}{$\nc=50$, $\nho=18$, $-1\le{}Z\le+3$} \\
$17$           & $\phantom{+}0$ & & $2.2(-1)$ & $8.1(-2)$ & $1.6(-1)$ & $2.1(-1)$ & $2.2(-1)$ & & $2.2(+1)$ & $5.1(+1)$ & $2.9(+0)$ & $1.8(+1)$ & $2.1(+1)$ \\
$17$           & $+1$           & & $3.5(-2)$ & $2.0(-3)$ & $2.6(-2)$ & $2.2(-2)$ & $3.6(-2)$ & & $1.2(+1)$ & $2.0(+0)$ & $3.0(+0)$ & $6.8(+0)$ & $1.2(+1)$ \\
$17$           & $+2$           & & $4.3(-4)$ & $8.7(-6)$ & $9.7(-4)$ & $1.5(-4)$ & $6.1(-4)$ & & $1.3(-1)$ & $4.9(-3)$ & $2.4(-1)$ & $2.4(-2)$ & $1.9(-1)$ \\
$18$           & $-1$           & & $2.8(+1)$ & $4.4(+1)$ & $1.1(+0)$ & $3.0(+1)$ & $2.6(+1)$ & & $-      $ & $-      $ & $-      $ & $-      $ & $-      $ \\
$18$           & $\phantom{+}0$ & & $2.1(+1)$ & $6.2(-1)$ & $4.8(+1)$ & $2.1(+1)$ & $2.0(+1)$ & & $5.6(+1)$ & $4.6(+1)$ & $2.9(+1)$ & $5.9(+1)$ & $5.5(+1)$ \\
$18$           & $+1$           & & $3.3(-2)$ & $1.3(-3)$ & $3.7(-1)$ & $4.9(-2)$ & $3.1(-2)$ & & $9.0(+0)$ & $1.2(+0)$ & $1.0(+1)$ & $1.5(+1)$ & $8.3(+0)$ \\
$18$           & $+2$           & & $7.5(-4)$ & $8.9(-6)$ & $8.3(-3)$ & $1.1(-3)$ & $4.3(-4)$ & & $2.3(-1)$ & $5.4(-3)$ & $2.4(+0)$ & $3.5(-1)$ & $1.3(-1)$ \\
$19$           & $-1$           & & $9.6(-1)$ & $1.1(+0)$ & $7.6(-2)$ & $3.2(-1)$ & $5.6(-1)$ & & $-      $ & $-      $ & $-      $ & $-      $ & $-      $ \\
$19$           & $\phantom{+}0$ & & $1.8(-1)$ & $3.3(-3)$ & $1.1(+0)$ & $5.3(-2)$ & $2.0(-1)$ & & $5.3(-6)$ & $6.9(-6)$ & $9.1(-6)$ & $3.9(-7)$ & $8.2(-1)$ \\
$19$           & $+1$           & & $5.6(-3)$ & $3.0(-6)$ & $4.7(-1)$ & $3.1(-3)$ & $7.5(-3)$ & & $1.7(+0)$ & $-      $ & $4.5(+1)$ & $8.1(-1)$ & $2.4(+0)$ \\
$20$           & $-1$           & & $1.9(-1)$ & $1.4(-1)$ & $2.0(-3)$ & $1.2(-1)$ & $1.4(-1)$ & & $-      $ & $-      $ & $-      $ & $-      $ & $-      $ \\
$21$           & $-1$           & & $1.8(-1)$ & $5.2(-2)$ & $1.7(-3)$ & $1.2(-1)$ & $1.3(-1)$ & & $-      $ & $-      $ & $-      $ & $-      $ & $-      $ \\
$22$           & $-1$           & & $1.8(-1)$ & $2.5(-2)$ & $1.7(-3)$ & $1.2(-1)$ & $1.3(-1)$ & & $-      $ & $-      $ & $-      $ & $-      $ & $-      $ \\
$23$           & $-1$           & & $1.8(-1)$ & $1.6(-2)$ & $1.8(-3)$ & $1.2(-1)$ & $1.4(-1)$ & & $-      $ & $-      $ & $-      $ & $-      $ & $-      $ \\
$24$           & $-1$           & & $1.9(-1)$ & $1.3(-2)$ & $1.8(-3)$ & $1.2(-1)$ & $1.4(-1)$ & & $-      $ & $-      $ & $-      $ & $-      $ & $-      $ \\
$25$           & $-1$           & & $2.0(-1)$ & $1.2(-2)$ & $1.8(-3)$ & $1.2(-1)$ & $1.4(-1)$ & & $-      $ & $-      $ & $-      $ & $-      $ & $-      $ \\
$26$           & $-1$           & & $2.0(-1)$ & $1.1(-2)$ & $1.9(-3)$ & $1.2(-1)$ & $1.4(-1)$ & & $-      $ & $-      $ & $-      $ & $-      $ & $-      $ \\
$27$           & $-1$           & & $2.1(-1)$ & $1.1(-2)$ & $1.9(-3)$ & $1.3(-1)$ & $1.4(-1)$ & & $-      $ & $-      $ & $-      $ & $-      $ & $-      $ \\
$28$           & $-1$           & & $2.2(-1)$ & $1.1(-2)$ & $2.0(-3)$ & $1.3(-1)$ & $1.5(-1)$ & & $-      $ & $-      $ & $-      $ & $-      $ & $-      $ \\
$29$           & $-1$           & & $2.2(-1)$ & $1.2(-2)$ & $2.0(-3)$ & $1.3(-1)$ & $1.5(-1)$ & & $-      $ & $-      $ & $-      $ & $-      $ & $-      $ \\
$30$           & $-1$           & & $2.3(-1)$ & $1.3(-2)$ & $2.1(-3)$ & $1.3(-1)$ & $1.5(-1)$ & & $-      $ & $-      $ & $-      $ & $-      $ & $-      $ \\
$31$           & $-1$           & & $2.4(-1)$ & $1.4(-2)$ & $2.2(-3)$ & $1.3(-1)$ & $1.6(-1)$ & & $-      $ & $-      $ & $-      $ & $-      $ & $-      $ \\
$32$           & $-1$           & & $2.5(-1)$ & $1.6(-2)$ & $2.2(-3)$ & $1.4(-1)$ & $1.7(-1)$ & & $-      $ & $-      $ & $-      $ & $-      $ & $-      $ \\
$33$           & $-1$           & & $2.6(-1)$ & $2.7(-2)$ & $2.3(-3)$ & $1.4(-1)$ & $1.8(-1)$ & & $-      $ & $-      $ & $-      $ & $-      $ & $-      $ \\
$34$           & $-1$           & & $2.7(-1)$ & $8.3(-2)$ & $2.4(-3)$ & $1.4(-1)$ & $2.2(-1)$ & & $-      $ & $-      $ & $-      $ & $-      $ & $-      $ \\
$35$           & $-1$           & & $2.8(-1)$ & $4.6(-1)$ & $2.4(-3)$ & $1.4(-1)$ & $5.0(-1)$ & & $-      $ & $-      $ & $-      $ & $-      $ & $-      $ \\
$36$           & $-1$           & & $4.5(+1)$ & $5.3(+1)$ & $4.2(+1)$ & $4.5(+1)$ & $4.8(+1)$ & & $-      $ & $-      $ & $-      $ & $-      $ & $-      $ \\
$36$           & $\phantom{+}0$ & & $4.0(-1)$ & $3.0(-2)$ & $5.6(+0)$ & $3.8(-1)$ & $8.3(-1)$ & & $5.0(-5)$ & $3.5(-5)$ & $6.2(-5)$ & $5.0(-5)$ & $5.2(-5)$ \\
\hline
\multicolumn{14}{c}{$\nc=96$, $\nho=24$, $-2\le{}Z\le+4$} \\
$48$           & $-2$           & & $2.1(+1)$ & $2.0(+1)$ & $2.3(+1)$ & $2.1(+1)$ & $2.1(+1)$ & & $-      $ & $-      $ & $-      $ & $-      $ & $-      $ \\
$48$           & $-1$           & & $5.7(+1)$ & $7.9(+1)$ & $2.4(+1)$ & $5.7(+1)$ & $5.7(+1)$ & & $-      $ & $-      $ & $-      $ & $-      $ & $-      $ \\
$48$           & $\phantom{+}0$ & & $2.2(+1)$ & $7.1(-1)$ & $5.0(+1)$ & $2.2(+1)$ & $2.2(+1)$ & & $4.4(+1)$ & $8.7(+1)$ & $1.3(+1)$ & $4.4(+1)$ & $4.4(+1)$ \\
$48$           & $+1$           & & $1.3(-1)$ & $1.0(-2)$ & $1.1(+0)$ & $1.3(-1)$ & $1.3(-1)$ & & $4.2(+1)$ & $1.2(+1)$ & $2.9(+1)$ & $4.2(+1)$ & $4.2(+1)$ \\
$48$           & $+2$           & & $2.0(-2)$ & $3.8(-4)$ & $2.1(-1)$ & $2.0(-2)$ & $2.0(-2)$ & & $1.4(+1)$ & $5.0(-1)$ & $5.6(+1)$ & $1.4(+1)$ & $1.4(+1)$ \\
\hline
\end{longtable}
\begin{list}{}{}
\item[$^{\mathrm{a}}$] N.R.: normal rates, i.e., the rates as discussed in Section \ref{sec:pahmodel}. $100\Gamma_\el{X}$ and $0.01\Gamma_\el{X}$: the normal rate for process X in/decreased by a factor of $100$. The effect of increasing a certain rate is the same as decreasing the rate of the reverse process. ea: electron attachment; em: photoelectric emission; diss,H: photodissociation with loss of H or H$_2$; add,H: addition of H or H$_2$.
\item[$^{\mathrm{b}}$] Only those states are shown whose abundance or emissivity contribution is at least $0.1\%$ for the normal rates.
\item[$^{\mathrm{c}}$] Percentage fraction of the entire PAH population in the disk.
\end{list}

\end{document}